\documentclass[aps,superscriptaddress,preprint,showkeys,nofootinbib,floatfix]{revtex4}

\usepackage{graphicx}  %
\usepackage{amsmath}
\usepackage{amsfonts}
\usepackage{bm}  %
\usepackage{xcolor}
\usepackage{hyperref}
\usepackage{ulem}
\hypersetup{
    colorlinks,
    linkcolor={red!50!black},
    citecolor={blue!50!black},
    urlcolor={blue!80!black}
}

\newcommand{\bea}{\begin{eqnarray}}
\newcommand{\eea}{\end{eqnarray}}
\newcommand{\beq}{\begin{equation}}
\newcommand{\eeq}{\end{equation}}
\newcommand{\bqa}{\begin{eqnarray}}
\newcommand{\eqa}{\end{eqnarray}}

\def\mqo2{{\!\!\!}}

\begin{document}

\title{ An effective-field-theory analysis of Efimov physics in
  heteronuclear mixtures of ultracold atomic gases }

\author{Bijaya Acharya}\email{bacharya@utk.edu}
\affiliation{Department of Physics and Astronomy, University of
  Tennessee, Knoxville, TN 37996, USA}

\author{Chen Ji}\email{ji@ectstar.eu}
\affiliation{ECT*, Villa Tambosi, 38123 Villazzano (Trento), Italy}
\affiliation{INFN-TIFPA, Trento Institute for Fundamental Physics and Applications, Trento, Italy}
\affiliation{TRIUMF, 4004 Wesbrook Mall, Vancouver, BC
  V6T 2A3, Canada}

\author{Lucas Platter}\email{lplatter@utk.edu}
\affiliation{Department of Physics and Astronomy, University of
  Tennessee, Knoxville, TN 37996, USA}
\affiliation{Physics Division,
  Oak Ridge National Laboratory, Oak Ridge, TN 37831, USA}
\date{\today}

\begin{abstract}
  We use an effective-field-theory framework to analyze the Efimov
  effect in heteronuclear three-body systems consisting of two species
  of atoms with a large interspecies scattering length.  In the
  leading-order description of this theory, various three-body
  observables in heteronuclear mixtures can be universally
  parameterized by one three-body parameter.  We present the
  next-to-leading corrections, which include the effects of the finite
  interspecies effective range and the finite intraspecies scattering
  length, to various three-body observables. We show that only one
  additional three-body parameter is required to render the theory
  predictive at this order. By including the effective range and
  intraspecies scattering length corrections, we derive a set of
  universal relations that connect the different Efimov features near
  the interspecies Feshbach resonance. Furthermore, we show that these
  relations can be interpreted in terms of the running of the
  three-body counterterms that naturally emerge from proper
  renormalization.  Finally, we make predictions for recombination
  observables of a number of atomic systems that are of experimental
  interest.
\end{abstract}

\keywords{few-body systems, Efimov effect, ultracold atoms}

\smallskip
\maketitle

\section{Introduction}
\label{sec:introduction}
Three-body systems of identical bosons display the Efimov
effect~\cite{Efimov70} when the interatomic scattering length is much
larger than the range of the underlying interaction. The key signature
of the Efimov effect is the discrete scaling of observables. For
example, in the limit of infinitely large scattering length, the
three-body bound state energies are in a geometric progression with a
common ratio $\lambda^2$, where the scaling factor $\lambda$ has the
value 22.694 for a system of three identical
bosons~\cite{Braaten:2004rn}. Multiple experiments with ultracold
atomic gases consisting of identical bosons have found signatures of
the Efimov effect by measuring loss rates that are driven by
three-body recombination effects in these systems
\cite{Kraemer:2006,Gross:2009,pollack:2009,Gross:2010}. The Efimov
effect also exists for systems of distinguishable particles with
different mass ratios.  This fact motivated for example the first
experimental measurement in heteronuclear atomic systems by the
Florence group~\cite{Barontini:2009} and, more recently, the experiments reported in 
Refs.~\cite{PhysRevLett.111.105301,PhysRevLett.113.240402,PhysRevLett.112.250404,Wacker:2016}.  Such systems
also exist in nuclear physics as neutron-rich halo nuclei, which are
weakly bound nuclei consisting of a tightly bound core and a small
number of valence neutrons~\cite{Platter:2009gz,Ji:2015jgm}.

The discrete scaling invariance and other scaling laws among Efimov
features in heteronuclear three-body systems have been calculated in
the zero-range limit by Helfrich {\it et al.}~\cite{Helfrich:2010yr}
using an effective-field-theory framework. Various potential models,
such as a Gaussian potential with a finite range~\cite{Blume:2014}, 
a minimal zero-range model~\cite{Zinner2015}, and Lennard-Jones
potentials (with van der Waals tails)~\cite{PhysRevLett.109.243201},
have also investigated the Efimov physics in heteronuclear
systems. The two latter theoretical works found better agreement with
experimental results for shallower Efimov states but stronger
discrepancies for deeper states. Such discrepancies are expected to be
associated with the finite effective ranges.  It is therefore
important to understand the range corrections in a systematic
framework.

Effective field theory (EFT) has been shown to be a convenient tool to
estimate uncertainties related to higher-order corrections in a
model-independent manner. EFTs are based on a systematic low-energy
expansion in a small parameter that is formed by a ratio of two
separated scales inherent to the problem at hand. In systems that
display the Efimov effect, this parameter is the ratio of the range of
the interaction to the scattering length.

The renormalization of the so-called {\it short-range} EFT at leading order (LO) 
was first worked out by Bedaque {\it et al.} in Ref.~\cite{Bedaque:1998km}. Since then, 
it has been used extensively to describe the zero-range limit of atoms
interacting through a large scattering length. Finite effective range
corrections were first considered within this framework in
Ref.~\cite{Hammer:2001gh}. However, nuclear systems with
a fixed scattering length were considered in this work. In
Refs.~\cite{Ji:2010su,Ji:2011qg}, the effective range, $r_0$, was included for
the case of variable scattering length and it was found that within
the EFT a second three-body datum is required for the approach to be
predictive at this order.

An additional complication arises in heteronuclear three-body systems
because there are two different scattering lengths.  Near the
interspecies Feshbach resonance, the two identical atoms interact with
each other through a smaller scattering length leading to deviations
from the scaling laws.

In this paper, we consider a three-body system of two identical bosons
(denoted by 2) that interact with one distinguishable atom (denoted by
1) via a large $s$-wave interspecies scattering length $a_{12}$ and a
small intraspecies scattering length $a_{22}$.  Such a system could
for example be prepared by choosing an appropriate Feshbach resonance
in the Lithium-Cesium system
\cite{PhysRevLett.113.240402,PhysRevLett.112.250404}~\footnote{The
  Feshbach resonances chosen in these specific experiments, however,
  also had a large Cs-Cs scattering length, $a_{22}$. Here we study
  systems with the scale hierarchy $r_0,\,a_{22}\ll a_{12}$.}. We
study low-energy processes that occur at a typical momentum
$k\sim {a_{12}}^{-1}$ by expressing all observables as simultaneous
expansions in $kr_0$ and $ka_{22}$, where $r_0$ is the interspecies effective range. At LO in this expansion, we
recover the results obtained by Helfrich {\it et
  al.}~\cite{Helfrich:2010yr}. We work to next-to-leading order (NLO)
where corrections linear in $r_0/a_{12}$ and $a_{22}/a_{12}$ enter~\footnote
{Although we assume that the intraspecies effective range, $r_{22}$, is of the same magnitude as $r_0$, 
it will enter at next-to-next-to leading order.}. We
show that these corrections can be interpreted in terms of the
renormalization group of the EFT~\cite{Ji:2015hha}. We then 
propose analytic formulas that account for the higher-order corrections in a simple manner.

\section{Effective field theory}
\label{sec:effect-field-theory}
In terms of the atomic fields, $\psi_1$ and $\psi_2$, and the
molecular fields, $d_{12}$ and $d_{22}$, the EFT Lagrangian in a
heteronuclear three-body system can be written as~\cite{Helfrich:2010yr}
\begin{align}
 \label{eq:lagrangian}
  \mathcal{L} \,=\, &\psi_1^\dagger\left(i\partial_t+\frac{\nabla^2}{2m_1}\right)\psi_1+
\psi_2^\dagger\left(i\partial_t+\frac{\nabla^2}{2m_2}\right)\psi_2 \nonumber\\
&- d_{12}^\dagger\left(i\partial_t+\frac{\nabla^2}{2 (m_1+m_2)}-\Delta_{12}\right)d_{12} 
 + \Delta_{22} d_{22}^\dagger d_{22} \nonumber\\
&-g_{12}\left(d_{12}^\dagger\psi_1\psi_2+\psi_1^\dagger\psi_2^\dagger d_{12}\right)
-\frac{g_{22}}{\sqrt{2}}\left(d_{22}^\dagger\psi_2\psi_2+\psi_2^\dagger\psi_2^\dagger d_{22}\right)
-h d_{12}^\dagger\psi_2^\dagger d_{12}\psi_2
\end{align}
where $g_{12}$ and $g_{22}$ are the two-body, and $h$, the three-body
coupling constants. The masses of the particles of type 1 and type 2
are denoted with $m_1$ and $m_2$, respectively. Since $a_{22}$ is of
natural size, the coupling constant between the identical atoms (which 
we will assume to be bosons), ${g_{22}}$, is treated perturbatively.
The bare propagator of the $d_{22}$ field then satisfies that
\begin{equation}
\frac{i}{\Delta_{22}} = i\frac{4\pi a_{22}}{m_2 g_{22}^2}\,.
\end{equation}
The interspecies scattering length $a_{12}$ is large such that atoms 1
and 2 form a two-body shallow virtual or bound molecular state. The
propagator that represents this state is obtained by non-perturbative
treatment of the coupling constant $g_{12}$. The renormalized parameters $g_{12}$ and $\Delta_{12}$ 
can then be related to the effective range parameters by matching to the effective range expansion~\cite{Beane:2000fi}. To simplify the
introduction of a finite effective range, we employ a dynamical
$d_{12}$ field in Eq.~\eqref{eq:lagrangian}. In order to preserve
the structure of the low-energy expansion and to extract the correction
strictly linear in the effective range we expand this propagator as
\begin{equation}
\mathcal{D}_{12}(p_0,{\bf p}) = \mathcal{D}_{12}^{(0)}(p_0,{\bf p})+\mathcal{D}_{12}^{(1)}(p_0,{\bf p})+\ldots,  
\end{equation}
and obtain the LO propagator,
\begin{equation}
  \label{eq:dimer_lo}
i\mathcal{D}_{12}^{(0)}(p_0,{\bf p})=-i\frac{2\pi}{\mu g_{12}^2}\,\frac{1}{-\gamma+\sqrt{-2\mu (p_0-\frac{p^2}{2(m_1+m_2)})-i\epsilon}},
\end{equation}
and the NLO propagator,
\begin{equation}
  \label{eq:dimer_nlo}
i\mathcal{D}_{12}^{(1)}(p_0,{\bf p})=
-i\frac{\pi r_0}{\mu g_{12}^2}\,\frac{\gamma+\sqrt{-2\mu (p_0-\frac{p^2}{2(m_1+m_2)})-i\epsilon}}{-\gamma+\sqrt{-2\mu (p_0-\frac{p^2}{2(m_1+m_2)})-i\epsilon}},
\end{equation}
where $\mu = m_1m_2/(m_1+m_2)$, $\mu_{AD} = m_2(m_1+m_2)/(m_1+2m_2)$,
and $\gamma$ denotes the interspecies binding momentum, which relates
to $a_{12}$ by $1/a_{12} = \gamma - r_0\gamma^2 /2$. At LO, we simply
have $\gamma=1/a_{12}$.

\begin{figure}[t]
\centerline{\includegraphics[width=\textwidth,angle=0,clip=true]{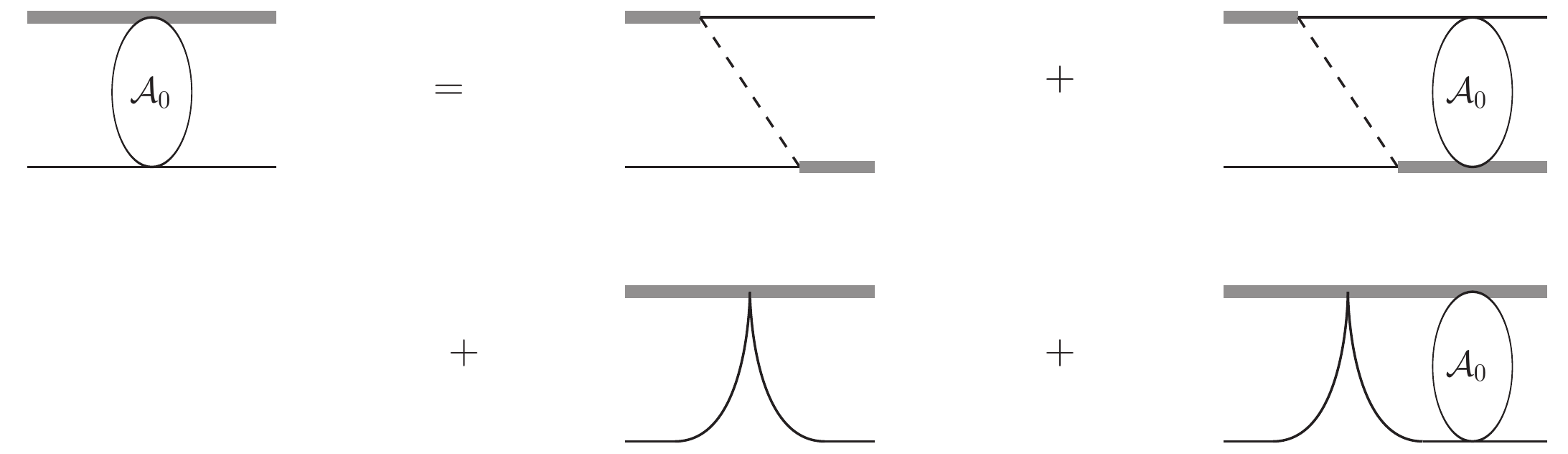}}
\caption{The STM equation for the LO scattering amplitude. The dashed
  and the solid lines represent the propagators of atoms 1 and 2
  respectively, and the thick gray line represents the dressed
  two-body propagator, $i\mathcal{D}_{12}^{(0)}$.}
\label{pic:stm}
\end{figure}

The atom-molecule ($\psi_1$-$d_{12}$) scattering amplitude given at LO by the Lagrangian
in Eq.~\eqref{eq:lagrangian} satisfies the Skorniakov--Ter-Martirosian
(STM) equation depicted in Fig.~\ref{pic:stm},
\begin{align}
\mathcal{A}_0(p,k;E)=&\frac{2\pi \gamma m_1}{\mu^2}\left[ K(p,k;E)+\frac{H^{(0)}(\Lambda)}{\Lambda^2}\right]
\nonumber\\
&+\frac{m_1}{\pi\mu}\int_0^{\Lambda}\hbox{d}q
\,q^2\left[K(p,q;E)+\frac{H^{(0)}(\Lambda)}{\Lambda^2} \right]
 \frac{\mathcal{A}_0(q,k;E)}{-\gamma+\sqrt{-2\mu(E-\frac{q^2}{2 \mu_{AD}})-i\epsilon}}. 
\label{eq:lo_amplitude}
\end{align}
where
$p$ and $k$ denote the relative momenta between the fields $\psi_1$ and $d_{12}$ in the three-body center of mass frame. The kernel function $K$ is defined as
\begin{equation}
K(p,q;E) = \frac{1}{2pq}\ln\frac{-2\mu E+p^2+q^2+2pq/(1+\delta)-i\epsilon}{-2\mu E+p^2+q^2-2pq/(1+\delta)-i\epsilon},
\end{equation}
where $\delta=m_1/m_2$ is the interspecies mass ratio
and the coupling constant $g_{12}$
has been eliminated by applying wave function
renormalization to the external legs of the amplitude, with a LO renormalization factor $Z_{12}=2\gamma/(\mu^2 g_{12}^2)$.

\begin{figure}[t]
\centerline{\includegraphics[width=10cm,angle=0,clip=true]{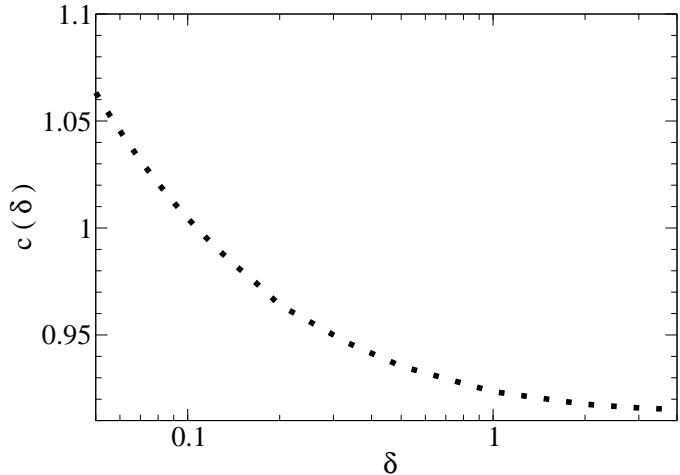}}
\caption{The numerical prefactor $c$ appearing in the LO three-body
  force $H_0(\Lambda)$ as a function of the mass ratio
  $\delta=m_1/m_2$.}
\label{pic:c-factor}
\end{figure}

The leading order three-body force parameter, $H^{(0)}(\Lambda)$,
that needs to be fixed by using one three-body datum as input, has
the analytic form
\begin{equation}
\label{eq:lo_cutoff_formula}
H^{(0)}(\Lambda)= \frac{2c(\delta)}{1+\delta}\frac{\sin[s_0\ln(\Lambda/\Lambda_\ast)+\arctan s_0]}{\sin[s_0\ln(\Lambda/\Lambda_\ast)-\arctan s_0]}~,
\end{equation}
and is a log-periodic function of $\Lambda$. The
three-body coupling $H^{(0)}$ is invariant under a discrete scaling
transformation by a scaling factor $\lambda = \exp(\pi/s_0)$, where
$s_0$ depends on the mass ratio $\delta$ (see the Appendix).  The factor $c(\delta)$, which
is obtained by matching Eq.~\eqref{eq:lo_cutoff_formula} to the
numerical value of $H^{(0)}(\Lambda)$, is required to render
Eq.~\eqref{eq:lo_amplitude} independent of the cutoff $\Lambda$. It is
shown as a function of $\delta$ in Fig.~\ref{pic:c-factor}.

At NLO, the three-body amplitude $\mathcal{A}$ can be modified as
\begin{equation}
\mathcal{A}(p,k;E)= (1+\gamma r_0) \mathcal{A}_0(p,k;E) + \mathcal{A}_1(p,k;E). 
\end{equation}

\begin{figure}[h]
\vspace{1cm}
\centerline{\includegraphics[width=0.9\textwidth,angle=0,clip=true]{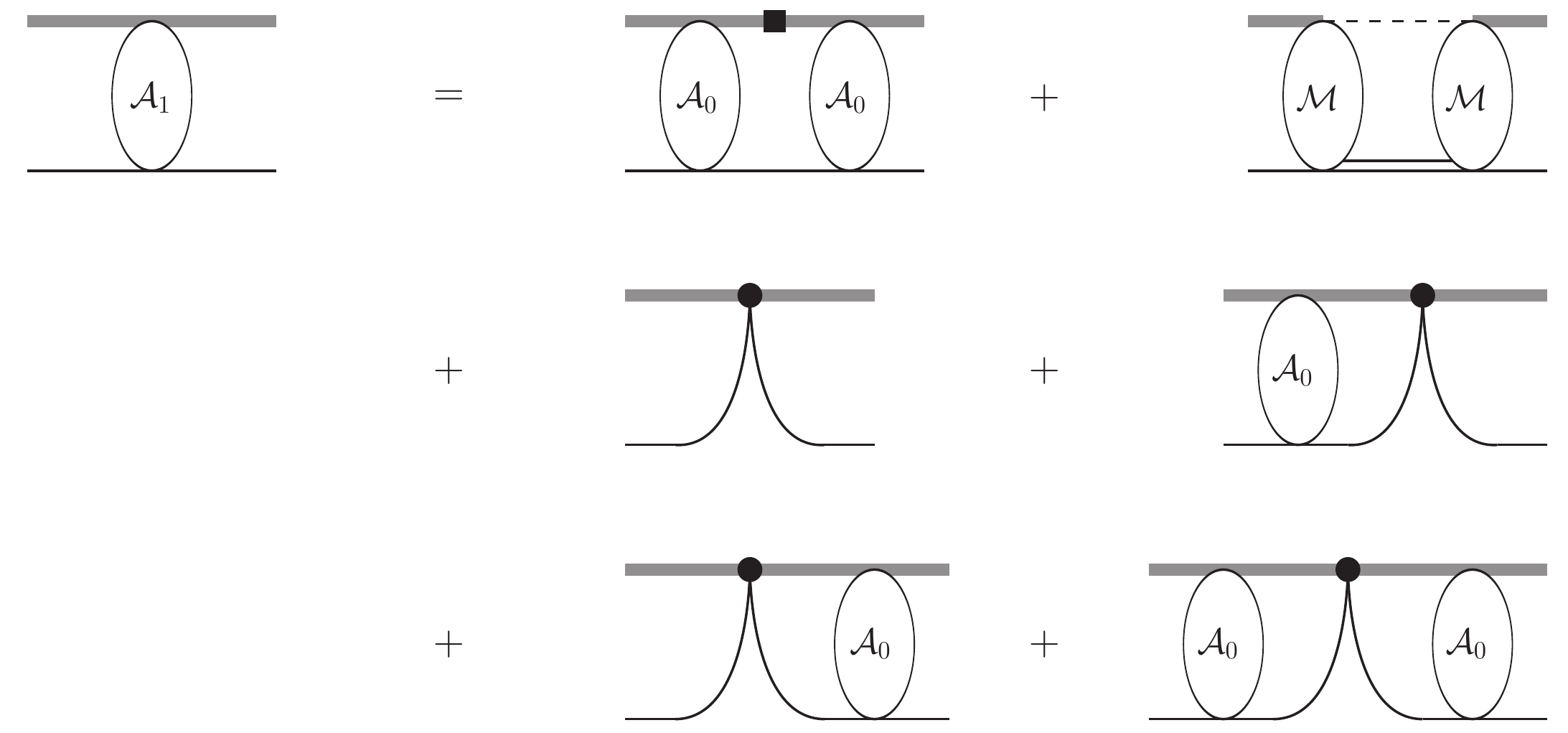}}
\vspace{1cm}
\centerline{\includegraphics[width=0.9\textwidth,angle=0,clip=true]{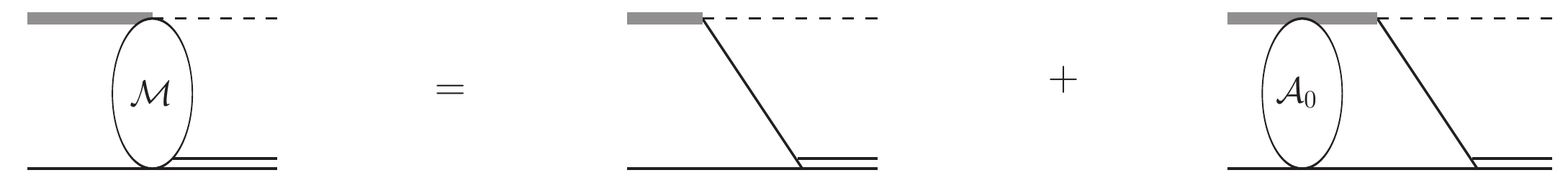}}
\caption{The upper equation shows the NLO scattering amplitude,
  $i\mathcal{A}_1$. The thick gray line with a black square represents
  the NLO dressed propagator, $i\mathcal{D}_{12}^{(1)}$, the double line
  represents the bare propagator of the $d_{22}$ field, and the
  three-atom vertex with circle represents the NLO three-body
  force. The lower equation defines the coupled-channel amplitude,
  which, up to constant factors, is equal to $\mathcal{M}$ defined in
  Eq.~\eqref{eq:coupled}. }
\label{pic:nlo}
\end{figure}

As shown in Fig.~\ref{pic:nlo}, the NLO amplitude
$\mathcal{A}_1(p,k;E)$ contains diagrams with one insertion of the NLO
propagator, $i\mathcal{D}_{12}^{(1)}$, and those with one insertion of the bare
propagator, $i/\Delta_{22}$. To absorb their linear and logarithmic
divergences, the diagrams with NLO three-body interactions are
introduced. These Feynman diagrams can be evaluated to give

\begin{align}
\label{eq:nlo_amplitude}
\mathcal{A}_1(p,k;E) = r_0 \frac{\mu}{4\pi^2\gamma}\int_0^{\Lambda} \hbox{d}q\,q^2 
&\frac{\gamma+\sqrt{-2\mu(E-\frac{q^2}{2 \mu_{AD}})-i\epsilon}}{-\gamma+\sqrt{-2\mu(E-\frac{q^2}{2 \mu_{AD}})-i\epsilon}}\mathcal{A}_0(p,q;E)\mathcal{A}_0(q,k;E)\nonumber\\
-a_{22}\frac{8\gamma m_2}{\mu^2}\int_0^{\Lambda} \hbox{d}q\,&q^2  \mathcal{M}(p,q;E)\,\mathcal{M}(k,q;E)\nonumber\\
			+ \frac{H^{(1)}(\Lambda)}{\Lambda^2}\frac{2\pi\gamma m_1}{\mu^2}&\left[1+\frac{\mu}{2\pi^2\gamma}\int_0^{\Lambda}\hbox{d}q\,q^2  \frac{\mathcal{A}_0(p,q;E)}{-\gamma+\sqrt{-2\mu(E-\frac{q^2}{2 \mu_{AD}})-i\epsilon}} \right] \nonumber\\
			\times&\left[1+\frac{\mu}{2\pi^2\gamma}\int_0^{\Lambda}\hbox{d}q\,q^2 \frac{\mathcal{A}_0(q,k;E)}{-\gamma+\sqrt{-2\mu(E-\frac{q^2}{2 \mu_{AD}})-i\epsilon}} \right],
\end{align}
where 
\begin{equation}
\mathcal{M}(p,k;E) = G(p,k;E)+\frac{\mu}{2\pi^2\gamma}\int_0^{\Lambda}\hbox{d}q\,q^2 
\frac{\mathcal{A}_0(p,q;E)}{-\gamma+\sqrt{-2\mu(E-\frac{q^2}{2 \mu_{AD}})-i\epsilon}} G(q,k;E),
\label{eq:coupled}
\end{equation}
with
\begin{equation}
G(p,k;E) = \frac{1}{2pk}\ln
\frac{-m_2E+p^2+\frac{m_2}{2\mu}k^2+pk-i\epsilon}
{-m_2E+p^2+\frac{m_2}{2\mu}k^2-pk-i\epsilon}~.
\end{equation}

The NLO three-body force has the form
\begin{equation}
H^{(1)}(\Lambda)  = \Lambda [r_0 h_{10}(\Lambda) + a_{22} \bar h_{10}(\Lambda)]
+\gamma [r_0 h_{11}(\Lambda) + a_{22}  \bar h_{11}(\Lambda)]~,
\end{equation}
where $h_{10}$, $h_{11}$, $\bar h_{10}$ and $\bar h_{11}$ can be
determined from experimental input.  The $\gamma$-independent pieces
of $H^{(1)}$, $h_{10}$ and $\bar h_{10}$, can be fixed by
renormalizing to the same observable that was used to reproduce the LO
three-body parameter. However, as discussed later in
Section~\ref{sec:univ-relat-renorm}, the $\gamma$-dependent parts of
$H^{(1)}$ requires tuning $h_{11}$ and $\bar h_{11}$ to fix one
additional three-body observable.  In practice, one can work at a
fixed value of $\Lambda$ for these counterterms. However, it is useful
to study their renormalization group evolution to ensure that the
regularization and renormalization have been carried out correctly and
consistently. We therefore analyze the renormalization-group flow of
the coupling constants, $h_{10}(\Lambda)$, $h_{11}(\Lambda)$,
$\bar h_{10}(\Lambda)$ and $\bar h_{11}(\Lambda)$ in the appendix.

\section{NLO corrections to recombination features}
\label{sec:nlo-corr-recomb}
The specific observables that are considered to be signatures of
Efimov physics are usually extracted from the so-called recombination
rate in experiments with ultracold atomic gases. Below we discuss the
NLO corrections to these features. Corresponding expressions for the
case of three identical bosons were derived in Ref.~\cite{Ji:2011qg}.

\subsection{Three-body binding energy}
The LO scattering amplitude $\mathcal{A}_0(p,k;E)$ at a given $\gamma$
has a series of discrete poles $E\rightarrow E_0^{(n)}(\gamma)$,
where $E_0^{(n)}$ are the energies of the $n$th three-body bound
states:
\begin{equation}
\mathcal{A}_0(p,k;E) = \frac{Z_0^{(n)}(p,k)}{E-E_0^{(n)}(\gamma)} + \mathcal{R}_0(p,k;E),
\end{equation}
where $Z^{(n)}_0(p,k)$ is the residue of the pole and
$\mathcal{R}_0(p,k;E)$ is the regular term around the pole
expansion. Due to NLO corrections, the pole position, the residue and
the regular term of $\mathcal{A}(p,k;E)$ are shifted by $E_1^{(n)}$,
$Z_1^{(n)}$, and $\mathcal{R}_1(p,k;E)$, respectively, i.e.,
\begin{equation}
\mathcal{A}(p,k;E) = \frac{Z_0^{(n)}(p,k)+Z_1^{(n)}(p,k)}{E-E_0^{(n)}(\gamma)-E_1^{(n)}(\gamma)} + \mathcal{R}_0(p,k;E) + \mathcal{R}_1(p,k;E),
\end{equation}
Matching terms linear in $r_0$ and $a_{22}$, we obtain
\begin{equation}
E_1^{(n)}(\gamma) = \frac{\displaystyle\lim_{E\rightarrow E_0^{(n)}}\left[E-E_0^{(n)}(\gamma)\right]^2\mathcal{A}_1(p,k;E)}
                 {\displaystyle\lim_{E\rightarrow E_0^{(n)}}\left[E-E_0^{(n)}(\gamma)\right]\mathcal{A}_0(p,k;E)}.
\end{equation}

\subsection{Three-body recombination rate}
\begin{figure}[t]
\centerline{\includegraphics[width=\textwidth,angle=0,clip=true]{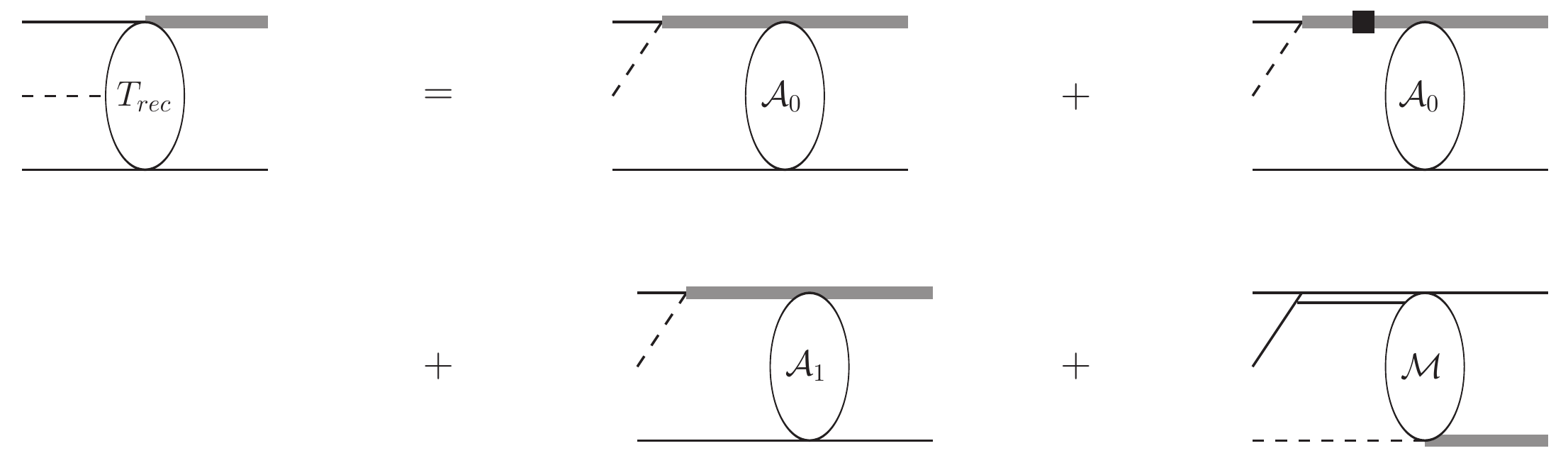}}
\caption{Amplitude for the recombination of free atoms into
  a shallow two-atom bound state and a residual atom. The first term on
  the right-hand side is the LO amplitude and the rest are NLO
  corrections.}
\label{pic:recmin}
\end{figure}

If $a_{12}>0$, three free atoms can recombine into a shallow
two-atom bound state and a residual atom. The energy released due to
the formation of the bound state is now converted to kinetic energy
and all three atoms leave the trap.  The Feynman diagrams that
contribute to the amplitude for this process, $iT_{rec}$, are shown in
Fig.~\ref{pic:recmin}.

The rate of change of the number densities of the atoms, $n_{1,2}$ due
to this recombination process is 
\begin{equation}
\frac{\mathrm{d}n_2}{\mathrm{d}t}=2\frac{\mathrm{d}n_1}{\mathrm{d}t}=-2\alpha n_1{n_2}^2,
\end{equation} 
where, up to NLO, the rate constant, $\alpha_s$ is given by 
\begin{equation}
\alpha_s=4\sqrt{\frac{\mu_{AD}^3}{\mu}}\,\frac{1}{\gamma^2}\,
\Bigg\lvert\mathcal{A}(0,\sqrt{\frac{\mu_{AD}}{\mu}}\gamma;0) 
+a_{22}\,4\pi\frac{\gamma^2}{\mu}\mathcal{M}(\sqrt{\frac{\mu_{AD}}{\mu}}\gamma,0;0) \Bigg\rvert^2.
\end{equation}

At LO, the rate constant has a minimum at $\gamma_0$ given by 
\begin{equation}
\mathcal{A}_0(0,\sqrt{\frac{\mu_{AD}}{\mu}}\gamma_0;0)=0.
\end{equation}
The NLO corrections shift the position of the recombination minimum to $\gamma_0+\Delta\gamma_0$, where $\Delta\gamma_0$ is given by the condition
\begin{equation}
 \frac{\mathrm{d}}{\mathrm{d}\gamma}\mathcal{A}_0(0,\sqrt{\frac{\mu_{AD}}{\mu}}\gamma;0)\Bigg\vert_{\gamma=\gamma_0}\Delta\gamma_0
 +\mathcal{A}_1(0,\sqrt{\frac{\mu_{AD}}{\mu}}\gamma_0;0)+a_{22}\,4\pi\frac{\gamma^2}{\mu}\mathcal{M}(\sqrt{\frac{\mu_{AD}}{\mu}}\gamma_0,0;0)=0.
\end{equation}

\subsection{Atom-molecule resonance}
The atom-molecule relaxation rate has a resonance when the a
three-body bound state crosses the atom-dimer threshold, i.e. when the
three-body bound-state energy coincides with the two-body bound-state
energy. The NLO shift in the resonance position, $\Delta\gamma_\ast$,
is given by the condition
\begin{equation}
 \Delta\gamma_\ast\displaystyle\lim_{\gamma\rightarrow\gamma_\ast}(\gamma-\gamma_\ast) \mathcal{A}_0(0,0;\frac{-\gamma^2}{2\mu}) 
 = \displaystyle\lim_{\gamma\rightarrow\gamma_\ast}(\gamma-\gamma_\ast)^2\mathcal{A}_1(0,0;\frac{-\gamma^2}{2\mu})
\end{equation}
where $\gamma_\ast$ is the resonance position at LO.

\subsection{Three-atom resonance}
\label{sec:three-atom-resonance}
Three-atom resonances occur when zero-energy three-body bound states
form, and result in maxima in the three-atom recombination rate. At
LO, this happens at $\gamma$ equals $\gamma_-<0$
when the Efimov curve crosses
the three-atom threshold.  The NLO correction to the resonance
position is given by
\begin{equation}
\Delta\gamma_- = -\frac{E_1^{(n)}(\gamma_-)}
{\qquad\frac{\mathrm{d}E_0^{(n)}(\gamma)}{\mathrm{d}\gamma}\Big\vert_{\gamma=\gamma_-}}.
\end{equation}

\section{Universal relations and renormalization group improvement}
\label{sec:univ-relat-renorm}
We use $a_i$, where $i$ runs over $0$, $\ast$ and $-$, to label the
values of $a_{12}$ associated with the signatures of the Efimov
effect. At LO, for which $a_i =1/\gamma_i$, the universal relations
between the various three-body observables can be summarized as
\begin{equation}
a_i^{(n)} = \lambda^n\theta_i\kappa_\ast^{-1},
\label{eq:lo_uv_rln}
\end{equation}
where $\kappa_\ast$ is the binding momentum of the $0$th state at the
unitary limit $\kappa_\ast\equiv\sqrt{2\mu\vert
  E_0^{(0)}(0)\vert}$. The values of $\lambda$ and $\theta_i$ for
various systems are listed in Table~\ref{tab:lo}.

\begin{table}
\centering
\begin{tabular*}{\textwidth}{@{\extracolsep{\fill}}cccccc}\hline\hline

System & $\delta$ & $\lambda$ & $\theta_0$ & $\theta_\ast$ & $\theta_-$ \\

\hline

$^6$Li-Cs-Cs & $4.511\times10^{-2}$ & $4.865$ & $0.6114$ & $3.388\times10^{-2}$ & $-1.349$ \\

$^7$Li-Cs-Cs & $5.263\times10^{-2}$ & $5.465$ & $0.5887$ & $3.392\times10^{-2}$ & $-1.376$ \\

$^6$Li-Rb-Rb & $6.897\times10^{-2}$ & $6.835$ & $0.5492$ & $3.367\times10^{-2}$ & $-1.436$ \\

$^7$Li-Rb-Rb & $8.046\times10^{-2}$ & $7.864$ & $0.5266$ & $3.328\times10^{-2}$ & $-1.477$ \\

$^{39}$K-Rb-Rb & $0.4483$ & $1.149\times10^2$ & $0.2247$ & $1.060\times10^{-2}$ & $-2.409$ \\

$^{40}$K-Rb-Rb & $0.4598$ & $1.227\times10^2$ & $0.2194$ & $1.014\times10^{-2}$ & $-2.430$ \\

$^{41}$K-Rb-Rb & $0.4713$ & $1.310\times10^2$ & $0.2142$ & $9.705\times10^{-3}$ & $-2.451$ \\

\hline\hline
  \end{tabular*}
\caption{LO universal Efimov parameters for different heteronuclear systems.\label{tab:lo}}
\end{table}

At NLO, we can write similar relations that express a desired
recombination feature as a sum of the LO universal relation and
shifts linear in $r_0$ and $a_{22}$,
\begin{equation}
  \label{eq:universal-relation1}
  a_i^{(n)}=\lambda^n\theta_i\kappa_\ast^{-1}+(J_i-n \sigma)r_0+(Y_i-n \bar{\sigma})a_{22}~,
\end{equation}
where $J_i$ and $Y_i$ are numbers that depend on the renormalization
condition chosen at NLO and $\sigma$ and $\bar{\sigma}$ are universal
numbers that depends only on the mass ratio in the heteronuclear
system. The quantities $J_i$ and $Y_i$ are not universal, however,
their differences, $(J_i-J_j)$ and $(Y_i-Y_j)$, are universal. 
To explicitly show the correlation 
between any three Efimov features in terms of universal numbers only,
we rewrite Eq.~\eqref{eq:universal-relation1} as
\begin{align}
\label{eq:ajm}
 a_j^{(m)} = &\,\frac{\lambda^m\theta_j-\lambda^l\theta_k}{\lambda^n\theta_i-\lambda^l\theta_k} 
 \left(a_i^{(n)}+\left[\left(J_j-J_i\right)-(m-n)\sigma\right]r_0
 +\left[\left(Y_j-Y_i\right) -(m-n)\bar\sigma\right]a_{22}  \right)\nonumber\\ 
  &+\frac{\lambda^m\theta_j-\lambda^n\theta_i}{\lambda^l\theta_k-\lambda^n\theta_i}\left(a_k^{(l)}+\left[\left(J_j-J_k\right)-(m-l)\sigma\right]r_0+\left[\left(Y_j-Y_k\right) -(m-l)\bar\sigma\right]a_{22}\right)
 \end{align}
which can be used to make predictions for all $j$ and $m$ using any two Efimov features, $a_i^{(n)}$ and $a_k^{(l)}$, as inputs. 
The values of the NLO universal
numbers for different systems are listed in
Table~\ref{tab:nlo}. Empirically, we find further correlations between these 
universal numbers. We obtain $J_0-J_-=\sigma/2$ and $Y_0-Y_-=\bar{\sigma}/2$ for all values of the mass ratio
$\delta$. 

\begin{table}
\centering
\begin{tabular*}{\textwidth}{@{\extracolsep{\fill}}ccccccc}\hline\hline

System & $\sigma=2(J_0-J_-)$ & $J_*-J_0$ & $\bar{\sigma}=2(Y_0-Y_-)$ & $Y_*-Y_0$ \\

\hline

$^6$Li-Cs-Cs & $0.693$ & $0.840$ & $0.141$ & $0.680$ \\

$^7$Li-Cs-Cs & $0.743$ & $0.828$ & $0.204$ & $0.821$ \\

$^6$Li-Rb-Rb & $0.840$ & $0.820$ & $0.367$ & 1.11 \\

$^7$Li-Rb-Rb & $0.904$ & $0.823$ & $0.502$ & 1.30 \\

$^{39}$K-Rb-Rb & 2.69 & 1.49 & 11.5 & 8.43 \\

$^{40}$K-Rb-Rb & 2.74 & 1.52 & 12.1 & 8.74 \\

$^{41}$K-Rb-Rb & 2.80 & 1.54 & 12.7 & 9.07 \\

\hline\hline
  \end{tabular*}
\caption{NLO universal Efimov parameters for different heteronuclear systems.\label{tab:nlo}}
\end{table}

A recent measurement of three-body recombination in an ultracold $^6$Li-Cs mixture determined the $^6$Li-Cs-Cs three-atom resonance in four consecutive Efimov states, 
whose positions $a_-^{(n)}$ are respectively $a_-^{(0)}=-350a_B$, $a_-^{(1)}=-1777a_B$, $a_-^{(2)}=-9210a_B$, and $a_-^{(3)}=-46635a_B$~\cite{Ulmanis2016}. 
The Cs-Cs scattering length $a_{22}$ near the $^6$Li-Cs Feshbach resonance is approximately $-1560a_B$ and varies slowly with $a_{12}$. 
Therefore, the two shallowest states satisfy the condition $|a_{22}|\ll |a_{12}|$, and are within the validity of the EFT description. Taking $a_-^{(2)}$, $a_-^{(3)}$, $a_{22}$ and $r_0\approx l_{vdw}=45a_B$ as inputs, 
we predict through Eq.~\eqref{eq:ajm} the Efimov features $a_0^{(2)}=4838a_B$, $a_0^{(3)}=22074a_B$, $a_*^{(4)}=5567a_B$, $a_*^{(5)}=28114a_B$, whose values satisfy that $r_0, |a_{22}|\ll |a_{12}|$ 
and deviate from the universal relations to $a_-^{(3)}$ by 11\%, 4.4\%, 2.3\%, and 1.4\% respectively. 
Although $a_-^{(0)}$ and $a_-^{(1)}$ in Ref.~\cite{Ulmanis2016} do not lie within the domain of validity of our $a_{22}/a_{12}$ expansion, we find that their values predicted by 
Eq.~\eqref{eq:ajm}, $-267a_B$ and $-1667a_B$, respectively, agree well with Ref.~\cite{Ulmanis2016}. Similar good agreement between the first order perturbative treatment of $a_{22}$ corrections 
with experiments even in the large $a_{22}/a_{12}$ regime is also seen in the correlations among the $a_-^{(n)}$ values observed in Ref.~\cite{PhysRevLett.113.240402}, and requires further investigation.

Similar to the LO universal parameter and discrete scaling factor, the NLO universal numbers can be also understood from the running of the three-body counterterms. 
The expressions for the three-body counterterms derived in the appendix 
are very similar in structure to the ones that have been derived for 
three identical bosons in Ref.~\cite{Ji:2011qg}. In particular, the
subleading three-body counterterms $h_{11}$ and $\bar h_{11}$ contain terms 
that indicate a logarithmic violation of the
leading order discrete scaling invariance. We can separate out these
terms by writing the full three-body force as
\begin{eqnarray}
\label{eq:H-RG-analytic}
\nonumber
  H(\Lambda)&=&
H_0(\Lambda)+h_{10}(\Lambda)\Lambda r_0 + \bar{h}_{10}(\Lambda)\Lambda a_{22}
\\
&&+ 
\left[\nu H'(\Lambda)\ln(\Lambda/Q_*) + \xi(\Lambda)\right]\gamma r_0
+
\left[\bar{\nu} H'(\Lambda)\ln(\Lambda/Q_*) + \bar\xi(\Lambda)\right] \gamma a_{22}~,
\end{eqnarray}
where $Q_*$ is the three-body parameter for the NLO renormalization,
which is determined by the additional three-body observable reproduced
at this order.  $H'(\Lambda)$ denotes the logarithmic derivative of
the LO three-body force, {\it i.e.},
$H'(\Lambda) \equiv d H(\Lambda)/(\Lambda d\Lambda)$. The dimensionless ratios $\nu$ and $\bar\nu$ 
are defined in the appendix. The redefined counterterms $\xi(\Lambda)$ and
$\bar\xi(\Lambda)$ in the above equation are periodic
functions of $\ln\Lambda$. Their explicit expressions are not necessary
for deriving the renormalization-group flow equation.

The term proportional to $\ln\Lambda/Q_*$ can be absorbed
into $H_0$ be defining a {\it running} Efimov parameter
\begin{equation}
\label{eq:running-efimov-parameter}
  \bar{\kappa_*}=(Q_*/\kappa_*)^{-\nu \gamma r_0 -\bar{\nu} \gamma a_{22}}  \kappa_*~.
\end{equation}

Now we can write down renormalization-group-improved universal
relations by elimination of $\kappa_*$ in Eq.~\eqref{eq:universal-relation1} in favor of the running
Efimov parameter $\bar{\kappa}_*$ defined in
Eq.~\eqref{eq:running-efimov-parameter}, i.e.,
\begin{equation}
  \label{eq:rg-improved-ur}
  a_{i,n}=
\lambda^n\theta_i(\lambda^n|\theta_i|)^{-(\nu r_0 +\bar{\nu} a_{22}) \kappa_*/(\lambda^n \theta_i)}\kappa_*^{-1}
+r_0 \tilde{J}_i+
  a_{22} \tilde{Y}_i~.
\end{equation}
Matching Eq.~\eqref{eq:universal-relation1} and Eq.~\eqref{eq:rg-improved-ur} yields
$\sigma=\nu\ln \lambda$ and $\bar{\sigma}=\bar{\nu} \ln \lambda$ and
\begin{eqnarray}
\nonumber
\tilde{J}_i&=&J_i+\nu \ln|\theta_i|,\\
\tilde{Y}_i&=&Y_i+\bar{\nu} \ln|\theta_i|.
\end{eqnarray}
The differences between the coefficients $\tilde{J}_j$ are universal
numbers and so are differences between the coefficients $\tilde{Y}_i$. 

In Refs.~\cite{Kievsky:2012ss,PhysRevA.88.032701} it was shown that a simple modification
of analytic expressions for zero-range observables can account in a
simple manner for a finite range corrections. In
Ref.~\cite{Ji:2015hha} it was shown that the underlying reason for
this simple approach is the slow logarithmic running of the modified
Efimov parameter shown above in
Eq.~\eqref{eq:running-efimov-parameter}.  In the heteronuclear system,
we can infer from the renormalization-group-improved universal relations above that the same strategy 
can account for higher-order corrections by modifying the analytic results presented in
Ref.~\cite{Helfrich:2010yr}. This will facilitate a simple inclusion
of the effects of deeply bound two-body states that have energies larger than
$1/(\mu r_0^2)$. 

For the recombination rate at positive scattering length, the authors
of Ref.~\cite{Helfrich:2010yr} found
\begin{equation}
  \label{eq:reco-analytical-lo-apos}
  \alpha_s=C(\delta)\frac{128\pi^2(4\pi-3\sqrt{3}) \left[\sin^2(s_0
    \ln(a_{12}/a_0))+\sinh^2\eta_*\right]}{\sinh^2(\pi
    s_0+\eta_*)+\cos^2(s_0\ln (a_{12}/a_0))}\frac{a_{12}^4}{m_1}~,
\end{equation}
where $C(\delta)$ is a mass dependent coefficient that has been
calculated in Ref.~\cite{Helfrich:2010yr}.  Following the prescription
laid out in Refs.~\cite{Kievsky:2012ss,PhysRevA.88.032701}, we
replace the $a_{12}^4$ factor in
Eq.~\eqref{eq:reco-analytical-lo-apos} with $\gamma^{-4}$ and
additionally introduce the parameter $\Gamma$ that shifts the
three-body parameter according to
$a_{i}^{-1}\rightarrow a_{i}^{-1}+\Gamma/a_{12}$. The parameter
$\Gamma$, which accounts for the corrections due to $r_0$ as well as those due to $a_{22}$, 
is different for each system and each observable $a_i$ and
can be fit to the data. Using these substitutions, we obtain
\begin{equation}
  \label{eq:reco-analytical-pheno}
  \alpha_s=C(\delta)\frac{128\pi^2(4\pi-3\sqrt{3}) \left[\sin^2(s_0
    \ln(a_{12}/a_0+\Gamma))+\sinh^2\eta_*\right]}{\sinh^2(\pi
    s_0+\eta_*)+\cos^2(s_0\ln (a_{12}/a_0+\Gamma))}\frac{1}{\gamma^4 m_1}
\end{equation}
Modifying the corresponding equation given in
Ref.~\cite{Helfrich:2010yr} for the rate of recombination into deeply bound two-body states
at positive scattering length leads to 
\begin{equation}
  \label{eq:reco-analytical-lo-aneg}
\alpha_d=C(\delta)\frac{\coth(\pi s_0)\cosh(\eta_*)\sinh(\eta_*)}
{\sinh^2(\pi s_0+\eta_*)+\cos^2(s_0 \ln(a_{12}/a_0+\Gamma))}\frac{1}{\gamma^4 m_1}~,
\end{equation}
where $\Gamma$ needs to have the same value as in
Eq.~\eqref{eq:reco-analytical-pheno}.  The total recombination rate
for positive scattering length is then given as the sum of
recombination into shallow and deeply bound two-body states, 
\begin{equation}
  \alpha=\alpha_s+\alpha_d~.
\end{equation}

The same conjecture can be made for negative scattering length and
the analytic expression derived in Ref.~\cite{Helfrich:2010yr} for
the rate of recombination into deeply bound two-body states leads to
\begin{equation}
  \label{eq:aneg-pheno}
  \alpha_d = \frac{C(\delta)}{2} \frac{128
    \pi^2(4\pi-3\sqrt{3})\coth(\pi s_0)\sin(2\eta_*)}
{\sin^2\left[s_0\ln(a_{12}/a_{-}+\Gamma')\right]+\sinh^2(\eta_*)}\frac{1}{\gamma^4 m_1} ~.
\end{equation}
where we used the parameter $\Gamma'$ to emphasize that it is different from the parameter $\Gamma$ used in the previous equations.

\section{Summary}
\label{sec:summary}
In this work we have calculated recombination features of the
heteronuclear three-body system at NLO in the short-range EFT
expansion. Specifically, we have considered systems in which the
interspecies scattering length is large compared to the van der Waals
length scale, and the effective ranges and the intraspecies scattering
length are of comparable size. At leading order in the EFT expansion,
only the interspecies scattering length and one three-body observable
are required within this approach for predictions. At NLO, a second
three-body observable is required in addition to the effective range
and the intraspecies scattering length. Our results give rise to
universal relations that can be used to predict recombination features
as a function of two-body scattering parameters and two three-body
observables. The parameters in these relations are universal and
depend only on the mass ratio in the heteronuclear system. We have
explicitly calculated these universal numbers for a number of physical
systems of interest. In particular, the ${}^6$Li-Cs-Cs seems to be
well suited to obtain experimental numbers to test our universal
relations. Alternatively, these relations could be tested using
few-body calculations with microscopic interactions as was done in
Ref.~\cite{Ji:2015hha}. In principle, we should be able to use the
ratios of three-atom threshold scattering lengths calculated by Blume
and Yan~\cite{Blume:2014} for such a comparison. We found 
that, while our results for these ratios are consistent with theirs, 
the numerical errors given in their work are too large to test our
NLO universal relations.

An extension of this research is to account for range corrections when
both the scattering lengths in the heteronuclear system become
simultaneously large. Work along these lines is under
development. However, generally, we expect universal relations that
account for finite range effects in other systems to look very similar
to the ones presented here. The most general case of course would be
to consider a system of three distinguishable particles with three
different scattering lengths.

A further important extension of our work is to apply this approach to
study systems at finite temperature in order to understand the
influence of temperature effects on the positions of recombination
features. For example, it was shown in
Ref. \cite{PhysRevA.91.063622} that finite range effects lead to
measurable temperature dependence of the recombination features. How
to include the effect of recombination into deeply bound two-body
states at next-to-leading order in effective field theory is also an open question.

\acknowledgments 

{This work was supported by the Office of Nuclear Physics,
  U.S.~Department of Energy under Contract No. DE-AC05-00OR22725, the
  National Science Foundation under Grant No. PHY-1516077 and by the
  Natural Sciences and Engineering Research Council (NSERC), and the
  National Research Council of Canada. LP thanks the ExtreMe Matter
  Institute EMMI at the GSI Helmholtz Centre for Heavy Ion Research
  for support and the Institute of Nuclear Physics at the TU Darmstadt
  for its hospitality during the completion of this work.}

\begin{appendix}
\section{Renormalization group evolution of the NLO three-body interaction}
\label{sec:renorm-group-evol}
Following Ref.~\cite{Ji:2011qg}, we can derive the asymptotic
expression for the LO amplitude $\mathcal{A}_0(p,k;E)$ for $p\gg k$,
$p\gg\sqrt{-2\mu E}$,
\begin{equation}
\mathcal{A}_0(p,k;E)\sim \widetilde{\mathcal{A}}_0(p) = \frac{z_0(p)}{p}+\gamma\frac{z_1(p)}{p^2}+\ldots,
\label{eq:asymptoticA}
\end{equation}
where the log-periodic functions $z_{0,1}$ are
\begin{equation}
 z_0(p)=\sin\left(s_0 \ln \frac{p}{\Lambda_\ast}\right),
\end{equation}
and
\begin{equation}
  z_1(p)=\frac{1}{\cos\phi}\lvert C_{-1} \rvert \sin\left(s_0 \ln \frac{p}{\Lambda_\ast}+\arg~C_{-1}\right).
\end{equation}
Here the constant $s_0$ and $C_{-1}$ are solved in a transcendental equation which satisfies that
\begin{equation}
\mathcal{I}(is_0) =1,
\end{equation}
and
\begin{equation}
C_{-1}=\frac{\mathcal{I}(is_0-1)}{1-\mathcal{I}(is_0-1)}.
\end{equation}
The function $\mathcal{I}(s)$ is defined as 
\begin{equation}
\mathcal{I}(s)=\frac{2\sin(\phi s)}{s\cos\left(\frac{\pi s}{2}\right)\sin(2\phi)}~,
\end{equation}
where $\phi=\arcsin(1/(1+\delta))$.

Using Eq.~\eqref{eq:asymptoticA} in Eq.~\eqref{eq:coupled}, we can
find a similar asymptotic form of $\mathcal{M}$(p,k;E) at $p\gg k$ and $p\gg\sqrt{-2\mu E}$,
\begin{equation}
\mathcal{M}(p,k;E)\sim \widetilde{\mathcal{M}}(p) = \frac{\sqrt{\mu\mu_{AD}}}{2\pi s_0 \gamma} \frac{\sinh\left(s_0\beta\right)}{\cosh\left(\frac{\pi s_0}{2}\right)} 
\left[\frac{1}{p}\,z_0\left(\frac{\rho p}{2}\right) + \frac{2\gamma}{\rho p^2}\,\bar z_1\left(\frac{\rho p}{2}\right)+\ldots\right],
\label{eq:asymptoticM}
\end{equation}
where $\rho = \sqrt{2m_2/\mu}$ and $\beta = \arcsin{\left(1/\rho\right)}$,
and
\begin{equation}
  \bar z_1(p)=\frac{1}{\cos\phi}\lvert D_{-1} \rvert \sin\left(s_0 \ln \frac{p}{\Lambda_\ast}+\arg~D_{-1}\right),
\end{equation}
with
\begin{equation}
 D_{-1}=\frac{s_0}{is_0-1}\,\frac{\sin\left(\beta[is_0-1]\right)}{\sinh\left(s_0\beta \right)}\,\frac{\cosh\left(\frac{\pi s_0}{2}\right)}{\cos\left(\frac{\pi}{2}[i s_0-1]\right)}\,\left(1+C_{-1}\right).
\end{equation}

Using Eqs.~\eqref{eq:asymptoticA} and \eqref{eq:asymptoticM} the
ultraviolet behavior of the integrals in Eq.~\eqref{eq:nlo_amplitude}
can be analyzed. There are linear and logarithmic divergences
proportional to both $r_0$ and $a_{22}$ which can be removed by
appropriate choices for the renormalization group evolution of the
counter terms $h_{10}(\Lambda)$, $h_{11}(\Lambda)$,
$\bar h_{10}(\Lambda)$ and $\bar h_{11}(\Lambda)$. The expressions for
the running of $h_{10}(\Lambda)$ and $\bar h_{10}(\Lambda)$ are,

\begin{equation}
\label{eq:h10}
  h_{10}(\Lambda)= -\frac{\pi(1+s_0^2)}{8}\sin(2\phi)\cos\phi\frac{(1+4s_0^2)^\frac{1}{2}-\cos\left(2s_0 \ln \frac{\Lambda}{\Lambda_\ast}-\arctan(2s_0)\right)}
  {(1+4{s_0}^2)^\frac{1}{2}\sin^2\left(s_0 \ln \frac{\Lambda}{\Lambda_\ast}-\arctan s_0\right)}
\end{equation}

and
\begin{equation}
\label{eq:bar_h10}
  \bar h_{10}(\Lambda)= \frac{2\pi}{\delta}\frac{1+s_0^2}{{s_0}^2}\frac{\sinh^2\left(s_0\beta\right)}{\cosh^2\left(\frac{\pi s_0}{2}\right)}\frac{(1+4{s_0}^2)^\frac{1}{2}
  -\cos\left(2s_0 \ln \frac{\rho\Lambda}{2\Lambda_\ast}-\arctan(2s_0)\right)}
  {(1+4{s_0}^2)^\frac{1}{2}\sin^2\left(s_0 \ln \frac{\Lambda}{\Lambda_\ast}-\arctan s_0\right)}.
\end{equation}

The value of $\Lambda_\ast$ in these equations is determined by the LO
renormalization condition and can be obtained from
Eq.~\eqref{eq:lo_cutoff_formula}. Eqs.~\eqref{eq:h10} and
\eqref{eq:bar_h10} are, therefore, predictive. 

\begin{figure}[t]
\centering
\includegraphics[width=0.49\textwidth,angle=0,clip=true]{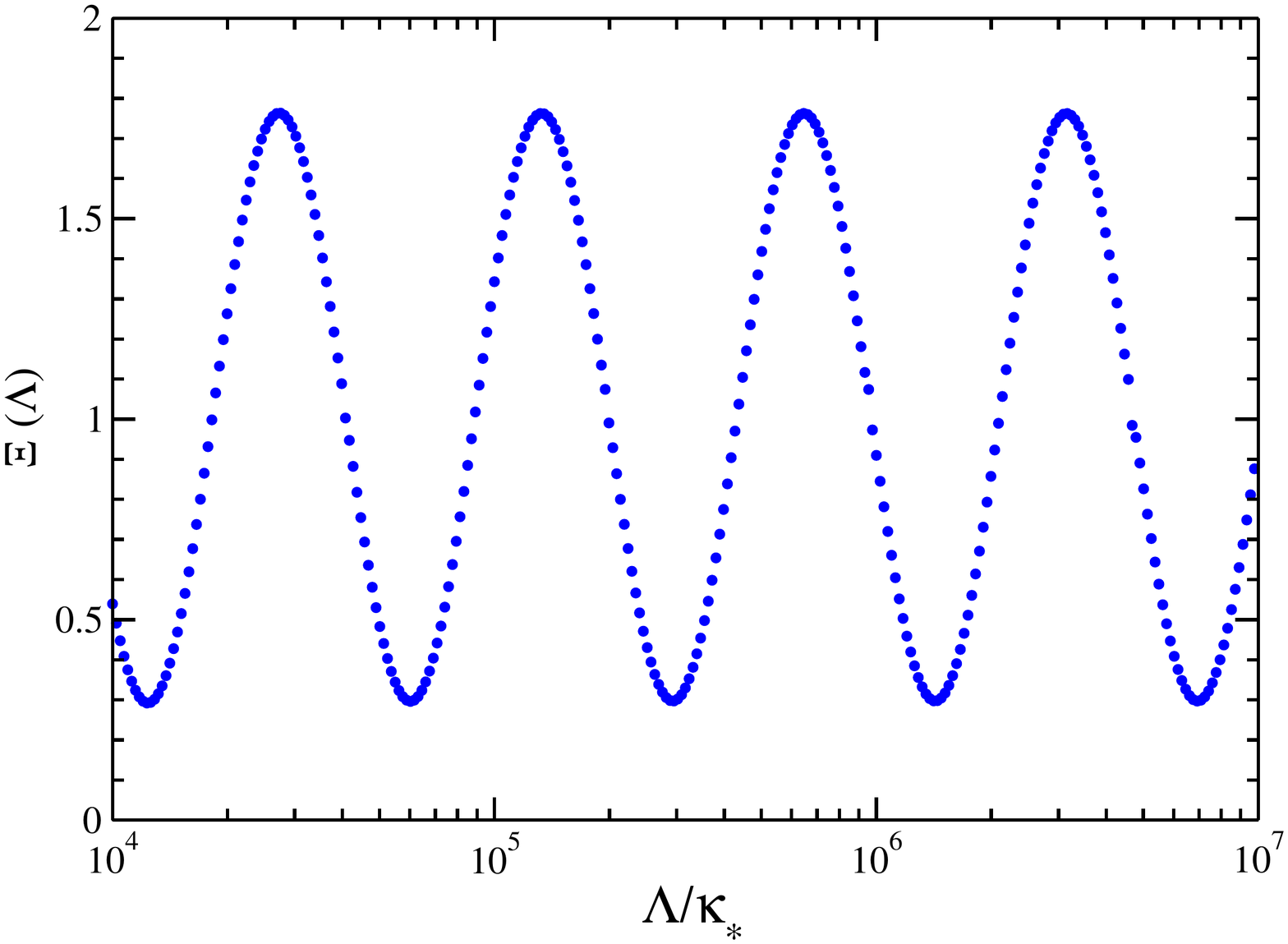}
\includegraphics[width=0.49\textwidth,angle=0,clip=true]{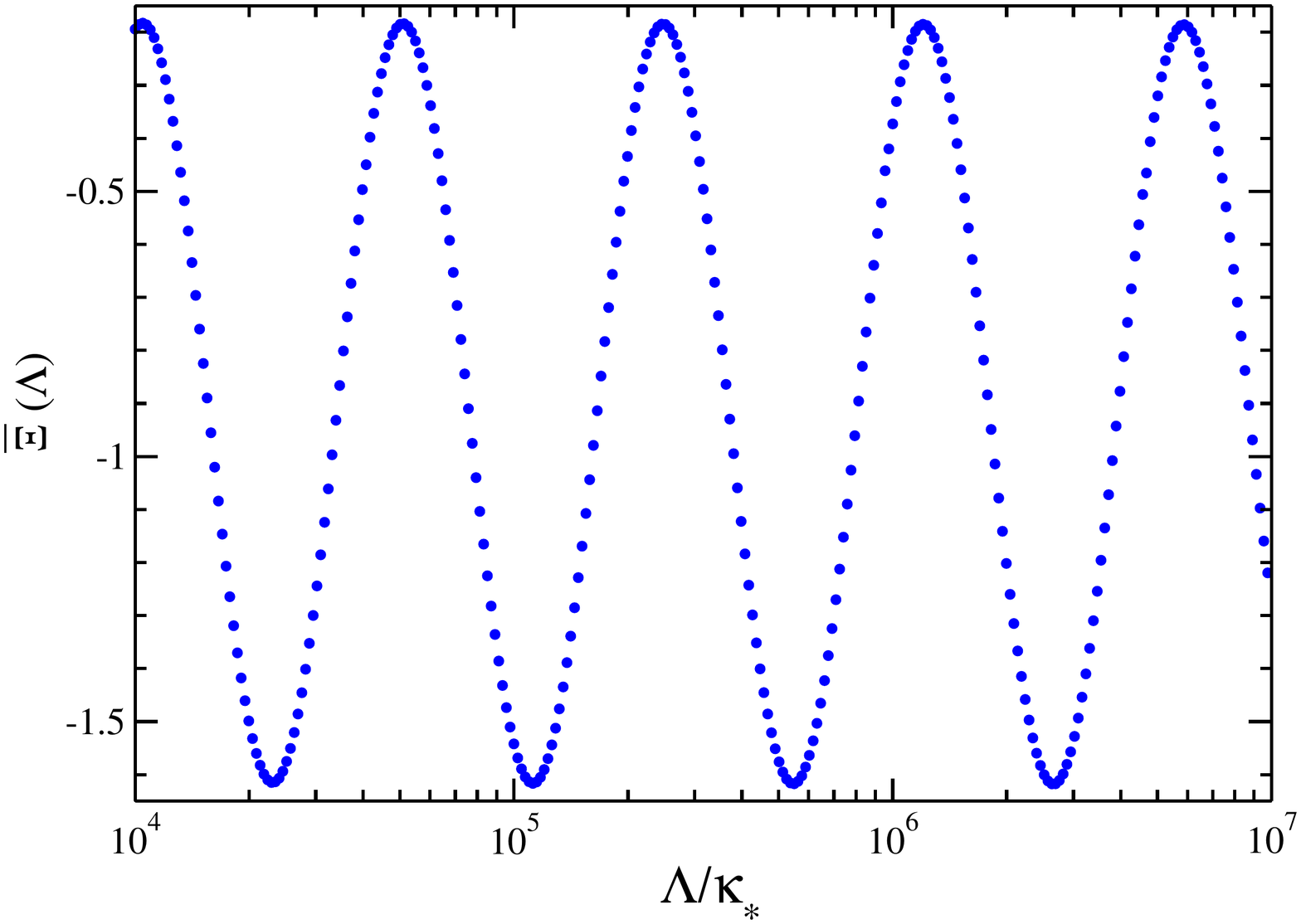}
\caption{The functions
  $ \Xi(\Lambda) \equiv \xi(\Lambda) \sin^2\left(s_0 \, \ln
    \frac{\Lambda}{\Lambda_\ast}-\arctan s_0\right)$ and
  $\bar\Xi \equiv \bar\xi(\Lambda) \sin^2\left(s_0 \, \ln
    \frac{\Lambda}{\Lambda_\ast}-\arctan s_0\right)$ for the
  $^6$Li-Cs-Cs system.  The renormalization conditions were
  $E_1^{(0)}(0)=0$ and $\Delta\gamma_0=0$. $Q_\ast$ was chosen to be
  equal to $\kappa_\ast$. Periodicity of these functions in
  $\ln\Lambda$ could be obtained by requiring both $d_R$ and
  $\bar d_R$ to have the value 0.981.}
\label{pic:nlo_running}
\end{figure}

Similarly, the expressions for the running of $h_{11}(\Lambda)$ and
$\bar h_{11}(\Lambda)$ are,
\begin{align}
  \label{eq:h11}
  h_{11}(\Lambda) = \,-d_K(\delta)\,\frac{\pi(1+s_0^2)}{4}\sin(2\phi)\frac{1+\lvert C_{-1}\rvert\cos\left(\arg~C_{-1}\right)}{\sin^2\left(s_0 \ln \frac{\Lambda}{\Lambda_\ast}-\arctan s_0\right)}\,\ln \frac{\Lambda}{Q_\ast} + \xi(\Lambda),
\end{align}

and
 \begin{align}
\label{eq:bar_h11}
 \bar h_{11}(\Lambda)= \,& \bar d_K(\delta)\,\frac{8\pi}{\delta\cos\phi}\frac{1+s_0^2}{s_0^2}\frac{\sinh^2\left(s_0\beta\right)}{\cosh^2\left(\frac{\pi s_0}{2}\right)} 
 \,\frac{1}{\rho} \, \frac{\lvert D_{-1}\rvert\cos\left(\arg D_{-1}\right)} {\sin^2\left(s_0 \, \ln \frac{\Lambda}{\Lambda_\ast}-\arctan s_0\right)}\,\ln\frac{\Lambda}{Q_*} \nonumber\\
 & + \bar\xi(\Lambda),
\end{align}

where $\xi(\Lambda)$ and $\bar\xi(\Lambda)$ are periodic functions of
$\ln\Lambda$, and $d_K(\delta)$ and $\bar d_K(\delta)$ are numerical
constants whose values are close to 1.  To make the arguments of the logarithms
dimensionless, we use the momentum scale $Q_\ast$. The constants
$d_K(\delta)$ and $\bar d_K(\delta)$, which are independent of the
choice of $Q_\ast$, can be determined by numerically evaluating the
running of $h_{11}(\Lambda)$ and $\bar h_{11}(\Lambda)$ while
maintaining the renormalization group invariance of physical
observables, and then demanding that Eqs.~\eqref{eq:h11} and
\eqref{eq:bar_h11} yield log-periodic values for the functions
$\xi(\Lambda)$ and $\bar\xi(\Lambda)$. To illustrate, we plot these
functions for the $^6$Li-Cs-Cs system for a particular choice of
renormalization conditions in Fig.~\ref{pic:nlo_running}. We find numerically that $d_K(\delta) = \bar d_K(\delta)$
for all systems. This equality stems from the fact that the regularization and the renormalization schemes
for both the counterterms $h_{11}$ and $\bar h_{11}$ are the same.

Eq.~\eqref{eq:H-RG-analytic} can now be obtained from Eqs.~\eqref{eq:lo_cutoff_formula}, \eqref{eq:h11} and \eqref{eq:bar_h11} by defining 
\begin{equation}
\nu = \frac{\pi (1+s_0^2)^2}{8s_0^2} \cos\phi 
\left[1+\lvert C_{-1}\rvert\cos\left(\arg C_{-1}\right)\right] 
\frac{d_K(\delta)}{c(\delta)}~, 
\end{equation}

and
\begin{equation}
\bar\nu = -\frac{4\pi}{\delta\sin (2\phi)} \frac{\left(1+s_0^2\right)^2}{s_0^4}\frac{\sinh^2\left(s_0\beta\right)}{\cosh^2\left(\frac{\pi s_0}{2}\right)} 
\,\frac{1}{\rho}\, \lvert D_{-1}\rvert\cos\left(\arg D_{-1}\right)
\frac{\bar d_K(\delta)}{c(\delta)}~.
\end{equation}

\end{appendix}

\end{document}